# Graphs, Matrices, and the GraphBLAS: Seven Good Reasons

Jeremy Kepner[1,*], David Bader[2], Aydın Buluç[3,†],
John Gilbert[4], Timothy Mattson[5], Henning Meyerhenke[6]
[1]*Massachusetts Institute of Technology (MIT),* [2]*Georgia Institute of Technology (GaTech),*
[3]*Lawrence Berkeley National Laboratory (LBNL),* [4]*University of California Santa Barbara (UCSB),*
[5]*Intel Corporation,* [6]*Karlsruhe Institute of Technology (KIT)*

---

**Abstract**
The analysis of graphs has become increasingly important to a wide range of applications. Graph analysis presents a number of unique challenges in the areas of (1) software complexity, (2) data complexity, (3) security, (4) mathematical complexity, (5) theoretical analysis, (6) serial performance, and (7) parallel performance. Implementing graph algorithms using matrix-based approaches provides a number of promising solutions to these challenges. The GraphBLAS standard (`istc-bigdata.org/GraphBlas`) is being developed to bring the potential of matrix based graph algorithms to the broadest possible audience. The GraphBLAS mathematically defines a core set of matrix-based graph operations that can be used to implement a wide class of graph algorithms in a wide range of programming environments. This paper provides an introduction to the GraphBLAS and describes how the GraphBLAS can be used to address many of the challenges associated with analysis of graphs.

---

# 1 Introduction

Graphs are among the most important abstract data structures in computer science, and the algorithms that operate on them are critical to applications in bioinformatics, computer networks, and social media [Ediger et al 2010, Ediger et al 2011, Reidy et al 2012, Reidy & Bader 2013]. Graphs have been shown to be powerful tools for modeling complex problems because of their simplicity and generality [Staudt et al 2014a, Bergamini et al 2015]. For this reason, the field of graph algorithms has become one of the pillars of theoretical computer science, informing research in such diverse areas as combinatorial optimization, complexity theory, and topology. Graph algorithms have been adapted and implemented by the military, commercial industry, and researchers in academia, and have become essential in controlling the power grid, telephone systems, and, of course, computer networks.

Graph analysis presents a number of unique challenges in the areas of (1) software complexity, (2) data complexity, (3) security, (4) mathematical complexity, (5) theoretical analysis, (6) serial performance, and (7) parallel performance. Implementing graph algorithms using matrix-based approaches provides a number of promising solutions to these challenges.

The duality between the canonical representation of graphs as abstract collections of vertices and edges and a matrix representation has been a part of graph theory since its inception [Konig 1931, Konig 1936]. Matrix algebra has been recognized as a useful tool in graph theory for nearly as long (see [Harary 1969] and references therein, in particular [Sabadusi 1960, Weischel 1962, McAndrew 1963, Teh & Yap 1964, McAndrew 1965, Harary & Tauth 1966, Brualdi 1967]). However, matrices have not traditionally been used for practical computing with graphs, in part because a dense two-dimensional matrix is not an efficient representation of a sparse graph. With the growth of efficient data structures and algorithms for *sparse* matrices, it has become possible to develop a practical matrix-based approach to computation on large, sparse graphs.

---

*This material is based upon work supported by the National Science Foundation under Grant No. DMS-1312831. Any opinions, findings, and conclusions or recommendations expressed in this material are those of the author(s) and do not necessarily reflect the views of the National Science Foundation.
†The work of this author is supported by the Applied Mathematics program of the DOE Office of Advance Scientific Computing Research under contract number DE-AC02-05CH11231.

The GraphBLAS standard (`istc-bigdata.org/GraphBlas`) is being developed to bring the potential of matrix based graph algorithms to the broadest possible audience. The GraphBLAS mathematically defines a core set of matrix-based graph operations that can be used to implement a wide class of graph algorithms in a wide range of programming environments. This paper provides an introduction to the GraphBLAS and describes how the GraphBLAS can be used to address many of the challenges associated with the analysis of graphs.

# 2 GraphBLAS Standard

Data analytics and the closely related field of "big data" have emerged as leading research topics in both applied and theoretical computer science. While it has been shown that many problems can be addressed with a "map-reduce" style framework, as we move to the next level of sophistication in data analytics applications, graph algorithms that demand more than "map-reduce" will play an increasingly vital role. There are many ways to organize a collection of graph algorithms into a high level library to support data analytics [Staudt et al 2014b]. It is probably premature to standardize these graph APIs. The low level building blocks of graph algorithms, however, are well understood and we believe a suitable target for standardization. In particular, the representation of graphs as sparse matrices allows many graph algorithms to be composed from a modest set of linear algebra operations.

Our concern, however, is that as new researchers enter this expanding field of research, the linear algebraic foundation of this class of graph algorithms will fragment. Diversity at the level of the primitive building blocks of graph algorithms will not help advance the field of graph algorithms. It will hinder progress as groups create different overlapping variants of what should be common low level building blocks. Furthermore, diverse sets of primitives will complicate the ability of the vendor community to support this research with math tuned to the needs of these algorithms.

It is our view that the state of the art in constructing a large collection of graph algorithms in terms of linear algebraic operations is mature enough to support the emergence of a standard set of primitive building blocks. We believe it is critical that we move quickly so as new research groups enter this field we can prevent needless and ultimately damaging diversity at the level of the basic primitives supporting this research, thereby freeing up researchers to innovate and diversify at the level of higher level algorithms and graph analytics applications.

The deep connection between graphs and sparse matrices [Kepner & Gilbert 2011] has been recognized to such an extent that it has led to the development of the GraphBLAS standard for bringing these fields together [Mattson et al 2013, Mattson 2014, Gilbert 2014, Kepner & Gadepally 2014, Buluc et al 2014]. The core of this connection is the duality between the fundamental operation on graphs---Breadth First Search (BFS)---and the fundamental operation of matrices---matrix multiply (see Figure 1).

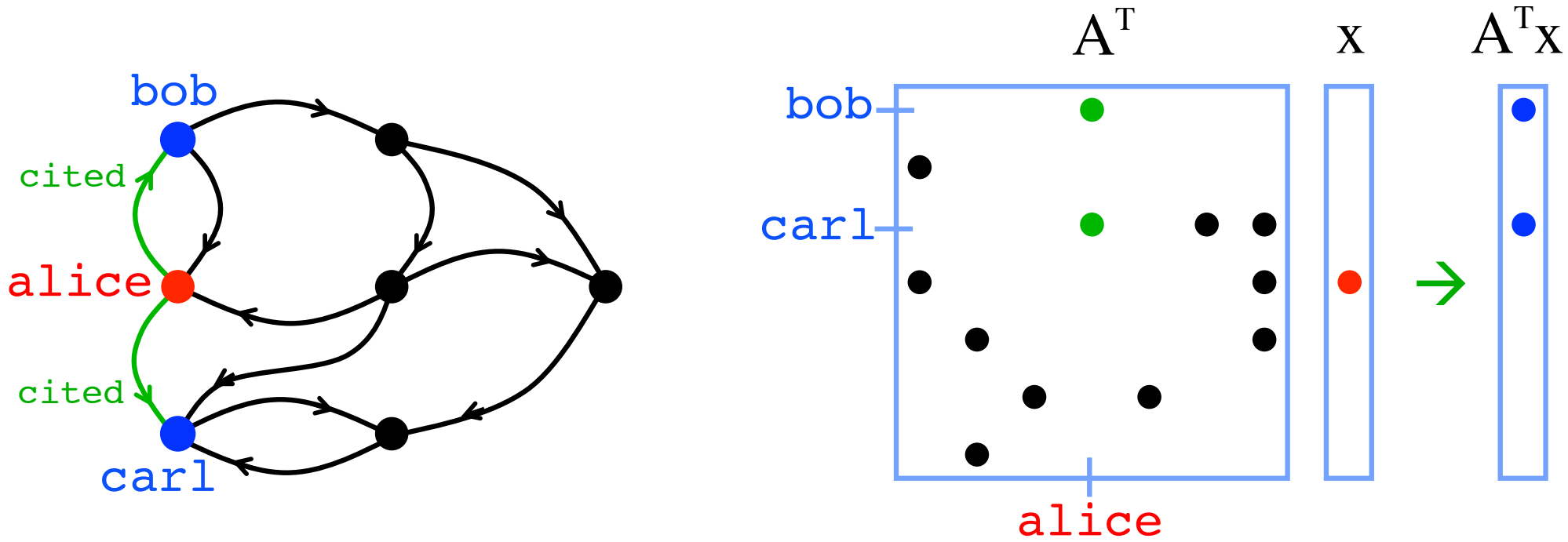

**Figure 1. Graph/Matrix Duality.** The fundamental operation of a graph is breadth first search (BFS) and is depicted on the left by the operation of starting at the node `alice` and traversing its edges to `bob` and `carl`. The identical operation is depicted on the right via multiplying the adjacency matrix representation of the graph by the vector with a single entry corresponding to the starting vertex `alice`.

The GraphBLAS define a narrow set of mathematical operations that have been found to be useful for implementing a wide range of graph operations. At the heart of the GraphBLAS are matrices. The matrices are usually sparse, which implies that the majority of the elements in the matrix are zero and are often not stored to make their implementation more efficient. Sparsity is independent of the GraphBLAS mathematics. All the mathematics defined in the GraphBLAS will work regardless of whether the underlying matrix is sparse or dense.

Graphs represent connections between vertices with edges. Matrices can represent a wide range of graphs using *adjacency* matrices or *incidence* matrices (defined below). Adjacency matrices are often easier to analyze while incidence matrices are often better for representing data. Fortunately, the two are easily connected by the fundamental mathematical operation of the GraphBLAS: matrix-matrix multiply. One of the great features of the GraphBLAS mathematics is that no matter what kind of graph or matrix is being used, the core operations remain the same. In other words, a very small number of matrix operations can be used to manipulate a very wide range of graphs.

## 2.1 Adjacency Matrix

A graph with N vertices and M edges can be represented by an N×N adjacency matrix $\mathbf{A}$, with rows and columns labeled by vertices. If $\mathbf{A}(v_1,v_2) = 1$, then there exists an edge going from vertex $v_1$ to vertex $v_2$. Likewise, if $\mathbf{A}(v_1,v_2) = 0$, then there is no edge from $v_1$ to $v_2$. The number of non-zero entries in $\mathbf{A}$ corresponds to the number of edges in the graph. Adjacency matrices have direction, which means that $\mathbf{A}(v_1,v_2)$ is not the same as $\mathbf{A}(v_2,v_1)$. Adjacency matrices can also have edge weights. If $\mathbf{A}(v_1,v_2) = w_{12}$, and $w_{12} \neq 0$, then the edge going from $v_1$ to $v_2$ is said to have weight $w_{12}$. Adjacency matrices provide a simple way to represent the connections between vertices in a graph between one set of vertices and another. Adjacency matrices are typically square and both out-vertices (rows) and the in-vertices (columns) are the same set of vertices. Adjacency matrices can be rectangular in which case the out-vertices (rows) and the in-vertices (columns) are different sets of vertices. This would occur, for example, in bipartite graphs. In summary, adjacency matrices can represent a wide range of graphs, which include any graph with any set of the following properties: directed, weighted, and/or bipartite.

## 2.2 Incidence Matrix

An M×N incidence, or edge matrix **E**, uses the rows to represent every edge in the graph and the columns represent every vertex. There are a number of conventions for denoting an edge in an incidence matrix. One such convention is to set $\mathbf{E}(i,v_1) = -1$ and $\mathbf{E}(i,v_2) = 1$ to indicate that edge i is a connection from $v_1$ to $v_2$. Incidence matrices are useful because they can easily represent multi-graphs, hyper-graphs, and multi-partite graphs. These complex graphs are difficult to capture with an adjacency matrix. A multi-graph has multiple edges between the same vertices. If there was another edge, j, from $v_1$ to $v_2$, this can be captured in an incidence matrix by setting $\mathbf{E}(j,v_1) = -1$ and $\mathbf{E}(j,v_2) = 1$. In a hyper-graph, one edge can go between more than two vertices. For example, to denote edge i has a connection from $v_1$ to $v_2$ and $v_3$ can be accomplished by also setting $\mathbf{E}(i,v_3) = 1$. Furthermore, $v_1$, $v_2$, and $v_3$ can be drawn from different classes of vertices and so **E** can be used to represent multi-partite graphs. Thus, an incidence matrix can be used to represent a graph with any set of the following graph properties: directed, weighted, multi-partite, multi-edge, and/or hyper-edge.

Using the conventions defined above for **E**, the adjacency matrix **A** and the incidence matrix **E** are linked by the formula $\mathbf{A} = |\mathbf{E}^\mathsf{T} < 0|\ |\mathbf{E} > 0|$. In other words, **A** is the cross-correlation of **E**. Any algorithm that can be written using **A** can also be written using **E** via the above formula. However, because **A** is a projection of **E**, information is always lost in constructing **A**, and there are algorithms that can be written using **E** that cannot be constructed using **A**.

## 2.3 Operations

The GraphBLAS consists of four core operations: matrix construction from triples (Sparse), extracting triples from a matrix (Find), element-wise addition (SpEWiseX), element-wise multiplication (SpEWiseX), and matrix products (SpGEMM)

$$\mathbf{A} = \mathbb{S}^{M\times N}(\mathbf{i},\mathbf{j},\mathbf{v}) \qquad (\mathbf{i},\mathbf{j},\mathbf{v}) = \mathbf{A} \qquad \mathbf{C} = \mathbf{A} \oplus \mathbf{B} \qquad \mathbf{C} = \mathbf{A} \otimes \mathbf{C} \qquad \mathbf{C} = \mathbf{A}\ \mathbf{B} = \mathbf{A} \oplus.\otimes \mathbf{B}$$

where:

$\mathbb{S}$ is a set of scalars (e.g., real numbers, complex numbers, integers, …);

**A**, **B**, **C** : $\mathbb{S}^{M\times N}$ are matrices of scalars;

**i**, **j**, and **v** are vectors corresponding to the row index, column index and value of the non-zero entries of a matrix;

$\oplus$ and $\otimes$ are user definable elment-wise "addition" and elment-wise "multiplication" operations;

$\oplus.\otimes$ denotes the matrix product using user defined $\oplus$ and $\otimes$ [Note: matrix products require that the number of rows of **A** are to equal the number of columns **B**].

Using these four operations, six additional operations can be constructed that form the GraphBLAS standard: Sparse, Find, Transpose, SpGEMM, SpRef, SpAsgn, SpEWiseX, Apply, and Reduce. These ten operations can then be used to build various utility functions and more complex graph operations and algorithms. The goal is to define a handful of matrix operations that can be implemented well and can enable a large class of graph algorithms. One possible implementation of these primitives can be found in the Combinatorial BLAS [Buluc and Gilbert 2011], which has been used to build higher level graph algorithms in the Knowledge Discovery Toolbox [Lugowski et al. 2012].

## 2.4 Composable Operations on Entire Graphs

Closure, associativity, distributivity, and commutativity are very powerful properties of the GraphBLAS and separate the GraphBLAS from standard graph libraries because these properties allow the GraphBLAS to be composable (i.e., you can re-order operations and know that you will get the same answer). Composability is what allows the GraphBLAS to implement a wide range of graph algorithms with just a few functions.

Let $\mathbf{A}$, $\mathbf{B}$, $\mathbf{C}$ : $\mathbb{S}^{N\times M}$, be matrices with scalar elements a = $\mathbf{A}$(i,j), b = $\mathbf{B}$(i,j), and c = $\mathbf{C}$(i,j) all drawn from the set of scalars $\mathbb{S}$. Associativity, distributivity, and commutativity of scalar operations translates into similar properties on matrix operations in the following manner.

*Closure* ensures that the result of combining two matrix representations of graphs using addition, elementwise multiplication, and/or matrix products will be another matrix representation of a graph. A unique aspect of the GraphBLAS is that the scalar addition ⊕ and scalar element-wise multiplication operations ⊗ are user definable. If these scalar operations have certain useful properties, then the corresponding matrix operations will also have those properties.

*Additive Commutativity* allows graphs to be swapped and combined via matrix element-wise addition without changing the result

$$a \oplus b = b \oplus a \quad \Rightarrow \quad \mathbf{A} \oplus \mathbf{B} = \mathbf{B} \oplus \mathbf{A}$$

where matrix element-wise addition is given by $\mathbf{C}(i, j) = \mathbf{A}(i, j) \oplus \mathbf{B}(i, j)$.

*Multiplicative Commutativity* allows graphs to be swapped, intersected, and scaled via matrix element-wise multiplication without changing the result

$$a \otimes b = b \otimes a \quad \Rightarrow \quad \mathbf{A} \otimes \mathbf{B} = \mathbf{B} \otimes \mathbf{A}$$

where matrix element-wise (Hadamard) multiplication is given by $\mathbf{C}(i, j) = \mathbf{A}(i, j) \oplus \mathbf{B}(i, j)$.

*Additive Associativity* allows graphs to be combined via matrix element-wise addition in any grouping without changing the result

$$(a \oplus b) \oplus c = a \oplus (b \oplus c) \quad \Rightarrow \quad (\mathbf{A} \oplus \mathbf{B}) \oplus \mathbf{C} = \mathbf{A} \oplus (\mathbf{B} \oplus \mathbf{C})$$

*Multiplicative Associativity* allows graphs to be intersected and scaled via matrix element-wise multiplication in any grouping without changing the result

$$(a \otimes b) \otimes c = a \otimes (b \otimes c) \quad \Rightarrow \quad (\mathbf{A} \otimes \mathbf{B}) \otimes \mathbf{C} = \mathbf{A} \otimes (\mathbf{B} \otimes \mathbf{C})$$

*Element-Wise Distributivity* allows graphs to be intersected and/or scaled and then combined or vice-versa without changing the result

$$a \otimes (b \oplus c) = (a \otimes b) \oplus (a \otimes c) \quad \Rightarrow \quad \mathbf{A} \otimes (\mathbf{B} \oplus \mathbf{C}) = (\mathbf{A} \otimes \mathbf{B}) \oplus (\mathbf{A} \otimes \mathbf{C})$$

*Matrix Multiply Distributivity* allows graphs to be transformed via matrix multiply and then combined or vice-versa without changing the result

$$a \otimes (b \oplus c) = (a \otimes b) \oplus (a \otimes c) \quad \Rightarrow \quad \mathbf{A}(\mathbf{B} \oplus \mathbf{C}) = (\mathbf{A}\mathbf{B}) \oplus (\mathbf{A}\mathbf{C})$$

where matrix multiply $\mathbf{C} = \mathbf{A}\mathbf{B}$ is given by

$$\mathbf{C}(i, j) = \oplus_k \mathbf{A}(i, k) \otimes \mathbf{B}(k, j)$$

for matrices $\mathbf{A}$ : $\mathbb{S}^{N\times M}$, $\mathbf{B}$ : $\mathbb{S}^{M\times L}$, and $\mathbf{C}$ : $\mathbb{S}^{N\times L}$.

*Matrix Multiply Associativity* is another implication of scalar distributivity and allows graphs to be transformed via matrix multiply in any grouping without changing the result

$$a \otimes (b \oplus c) = (a \otimes b) \oplus (a \otimes c) \quad \Rightarrow \quad (\mathbf{AB})\mathbf{C} = \mathbf{B}(\mathbf{AC})$$

# 3 Seven Good Reasons

Graph analysis presents a number of unique challenges in the areas of (1) software complexity, (2) data complexity, (3) security, (4) mathematical complexity, (5) theoretical analysis, (6) serial performance, and (7) parallel performance. Implementing graph algorithms using matrix-based approaches provides a number of promising solutions to these challenges.

## 3.1 Software Complexity

Nearly all negative attributes of software system (e.g., effort, schedule, testing, defects, security issues, documentation, …) increase with the size of the software. A typical graph library has hundreds of functions consisting of many thousands of lines of code [Boost, Lemon, JGraphT, GraphStream, Jung, …]. This high level of software complexity means that graph libraries require a lot of effort to implement. This effort increases when the functions are implemented to run in parallel. Furthermore, the large number of functions means that there is very little opportunity for hardware optimization. The effort of implementing a graph function in hardware is significantly greater and requires that a small number of critical functions be identified than can enable a large class of applications.

One approach to addressing software complexity is to use a set of composable functions that allow a large number of graph algorithms to be built with a small number of building blocks. Sparse matrix operations are one composable approach to building graph algorithms. Many graph algorithms can be represented with matrix algebra using just a handful of functions [Kepner & Gilbert 2011].

## 3.2 Data Complexity

Most graph libraries or graph databases define specific data structures for holding different types of graphs. These graph tools often begin their development with the simplest of graphs (undirected/unweighted) and create a vertex-oriented data structure that works well for this type of graph for the specific tool. As the tool matures, more complex graphs are addressed by either evolving the initial graph data structure or adding new data structures. This cumulative approach requires that each graph function be revisited each time the core data structure is changed. Furthermore, since the starting point is often a data structure designed for the simplest possible graphs, it can be difficult to evolve this to accommodate the most complex graphs that are often seen in the real world.

The GraphBLAS provide one approach to addressing this challenge by defining a primary data structure in the GraphBLAS that isn't a graph. The primary GraphBLAS data structure is a sparse matrix. Sparse matrices can be used to efficiently represent both simplest and complex graphs (see sections 2.1 and 2.2). The GraphBLAS functions are defined on sparse matrices and thus work regardless of the complexity of the graph that is being represented. Furthermore, sparse matrices have been demonstrated to work well in a wide range of programming environments and fit in naturally with modern key/value databases [Kepner et al 2013].

## 3.3 Security

Security is becoming increasingly important in wide range of applications, and security considerations are playing an increasing role in algorithm design. Executing algorithms on encrypted data requires modifying existing algorithms to include various cryptographic techniques such homomorphic encryption, fully homomorphic encryption, multi-party computation, deterministic encryption, and order preserving encryption. A focus of this work has been adapting scalar addition $\oplus$ and scalar $\otimes$ multiplication to work with encrypted data, which naturally lead to matrix operations working on encrypted data [Erkin 2010, Kepner et al 2014].

## 3.4 Mathematical Complexity

Standard graph libraries provide hundreds of functions, but still often require the user to write long programs to take advantage of these functions. This is often due to the wide variety of different data structures that exist in a standard graph library and the fact that graph library functions are strongly order dependent. The GraphBLAS provides a single mathematically defined object that is mathematically closed (i.e., produces the same class of outputs as inputs) under the aforementioned operations. This eliminates the requirement to write code to convert between different representations of a graph. That said, the conversion between a sparse matrix and a triples representation is explicitly supported so that GraphBLAS users can work the data as they please. In addition to being mathematically closed, the GraphBLAS operations are composable, which means that operations can be reordered without changing the results. As a result, it is possible to implement complex analytics with ~50x less effort than other approaches [Kepner et al 2011].

## 3.5 Theoretical Analysis

While matrix-based graph approaches have been around since the inception of graph theory, these approaches were less widely used for graph algorithm analysis. Interest in this area was dramatically increased by the advent of the legendary Google PageRank algorithm [Brin & Page 1998] that exploited the first eigenvector of the graph adjacency matrix. Since the development of PageRank, graph algorithm theorists have found matrix based approaches to graph analysis are highly productive [Madry et al 2010, Madry 2011, Madry et al 2011a, Madry et al 2011b, Madry et al 2013, Dodson et al 2014].

## 3.6 Serial Processing Performance

On conventional processors, there is a large gap between the performance of traditional dense matrix computations and sparse/graph computations (see Figure 1). This is partly because sparse/graph computations are typically memory bound, while dense matrix computations are compute bound. Memory performance improvements have been, and will most likely continue to be, slower than processor performance improvements. However, a part of the problem is that sparse/graph computations do rarely achieve their potential performance. While DGEMM can achieve >80% of peak flops, very few implementations of sparse/graph operations utilize 80% or more of the available system memory bandwidth (which is their corresponding upper bound). We know of only one existing effort where the "roofline" [Williams et al 2009] of a graph computation has been analyzed, which was the case for breadth-first search using sparse matrix algebra [Lugowski et al 2014]. The high actual to peak performance ratio of dense matrix computations is in large part due to the existence of the Basic Linear Algebra Subprograms (BLAS) standard [Lawson et al 1979]. The BLAS provide a concise set of functions that allow software developers to write hardware independent code and hardware designers to target their efforts on a small number of functions that impact a wide range of programs. It is the goal of the GraphBLAS to provide the same performance benefit to graph algorithms.

## 3.7 Parallel Processing Performance

Parallel graph algorithms are notoriously difficult to implement and optimize [Ediger et al 2012, Ediger & Bader 2013, Meyerhenke 2013 McLaughlin & Bader 2014a, McLaughlin & Bader 2014b, McLaughlin et al 2014, Staudt & Meyerhenke 2015, Meyerhenke et al 2015]. A matrix-based approach to graph algorithms allows the graph algorithms community to leverage the decades of work in creating optimized parallel algorithms for matrix computations. Even still, the inherently high communication to computation ratios found in graph algorithms mean that even the best algorithms will result in parallel efficiencies that decreases as the number of processors P is increased by a factor of $P^{1/2}$ (see Figure 2) [Buluc & Gilbert 2012]. Recent work on communication-avoiding algorithms, and their applications to graph computations [Ballard et al 2013, Solomonik et al 2013], might defer but not completely eliminate the parallel scalability bottleneck. Consequently, novel hardware architectures will also be required. The GraphBLAS simplifies this hardware design challenge by providing a clear target for system designers. In addition, because the GraphBLAS deals with graph computations in aggregate (instead of individually), it reduces the computational challenge to one of providing high bandwidth (instead of high bandwidth *and* low latency).

.

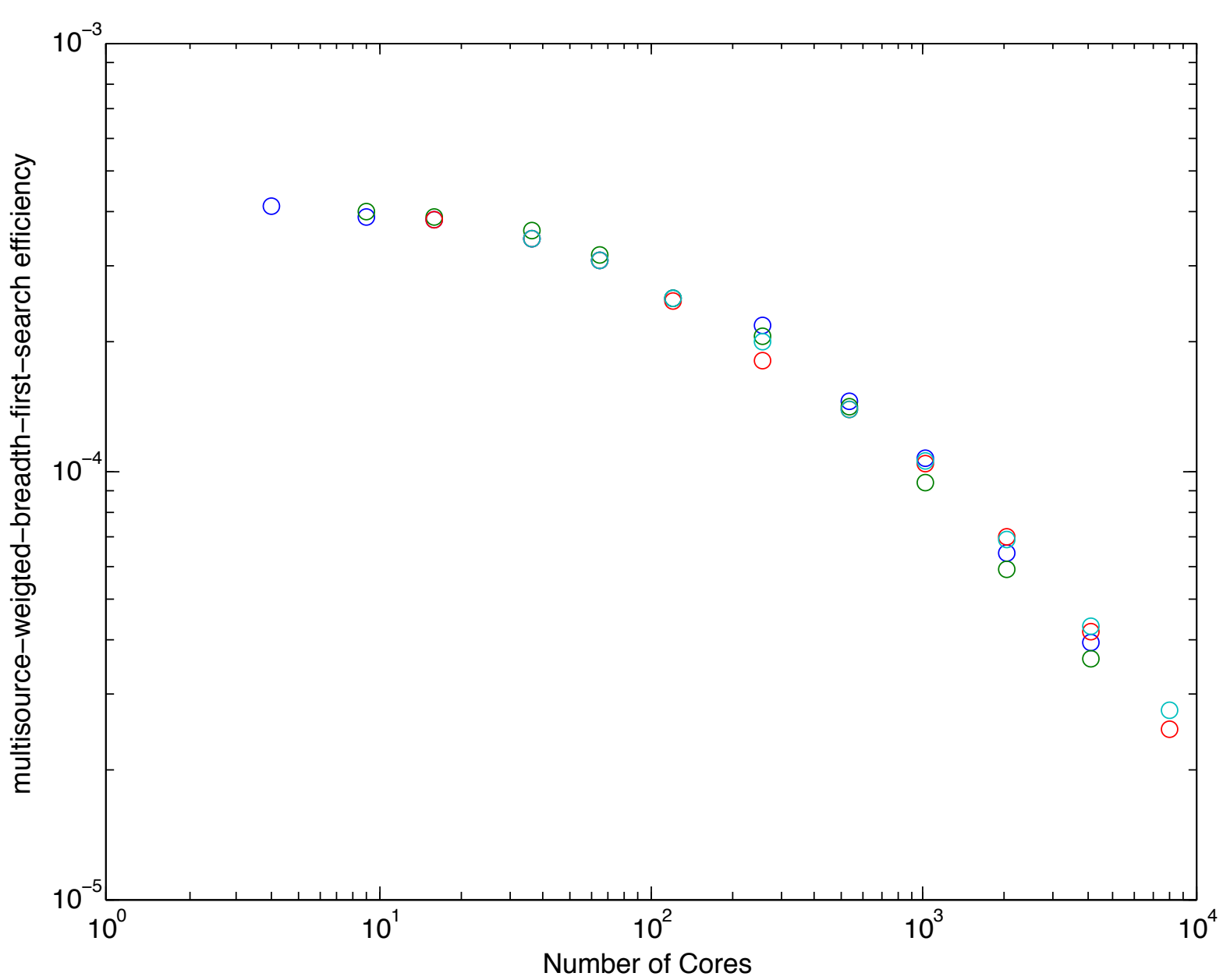

**Figure 2. Graph Computation Efficiency.** [Buluc & Gilber 2012] Performance of multi-source-weighted-breadth-first-search (i.e., sparse matrix-matrix multiply) on power-law graph on a Cray supercomputer. For the largest graph (Scale-24 ⇒ 16 million vertices and 268 million edges) a speedup of 20 is achieved even though the number of processors is increased by over 200.

# 4 Summary

The analysis of graphs have become increasingly important to a wide range of applications. Graph analysis presents a number of unique challenges in the areas of (1) software complexity, (2) data complexity, (3) security, (4) mathematical complexity, (5) theoretical analysis, (6) serial performance, and (7) parallel performance. Implementing graph algorithms using matrix-based approaches provides a number of promising solutions to these challenges. The GraphBLAS standard (`istc-bigdata.org/GraphBlas`) is being developed to bring the potential of matrix based graph algorithms to the broadest possible audience. The GraphBLAS mathematically defines a core set of matrix-based graph operations that can be used to implement a wide class of graph algorithms in a wide range of programming environments. This paper provides an introduction to the GraphBLAS and describes how the GraphBLAS can be used to address many of the challenges associated with analysis of graphs.